\documentclass[a4paper,11pt]{article}
\usepackage{wrapfig}
\usepackage{pos}
\usepackage{lipsum} 
\usepackage[mathlines]{lineno}

\title{Measurement of the Cosmic Neutrino Flux from the Southern Sky using 10 years of IceCube Starting Track Events}

\ShortTitle{Diffuse Flux Measurement using Starting Tracks}

\author{The IceCube Collaboration \\{\normalsize \normalfont(a complete list of authors can be found at the end of the proceedings)}\\}

\emailAdd{msilva@icecube.wisc.edu}
\emailAdd{sarah.mancina@icecube.wisc.edu}
\emailAdd{josborn@icecube.wisc.edu}

\abstract{

The measurement of a diffuse astrophysical neutrino flux using starting track events marks the first time IceCube has observed and subsequently measured the astrophysical diffuse flux using a dataset composed primarily of starting track events. Starting tracks combine an excellent angular and energy resolution. This enables us to take advantage of the self-veto effect in the southern sky reducing the atmospheric neutrino rate allowing us to detect astrophysical neutrinos to energies well below 100 TeV. We measure the astrophysical flux as $\phi^\mathrm{per-flavor}_{\mathrm{Astro}}=1.68^{+0.19}_{-0.22}$(at 100 TeV) and $\gamma_{\mathrm{Astro}} = 2.58^{+0.10}_{-0.09}$ assuming a single power law flux. The astrophysical flux $90\%$ sensitive energy range is 3 TeV to 500 TeV, extending IceCube's reach to the low energy astrophysical flux by an order of magnitude.
A brief summary of tests performed to search for neutrinos from the galactic plane using this dataset is also provided. With this sample, we did not find statistically significant evidence for emission from the galactic plane. We then tested the impact of these galactic plane neutrinos on the isotropic diffuse flux, with at most 10\% effect on the overall normalization and negligible impact to the spectral index.

\vspace{4mm}
{\bfseries Corresponding authors:}
Manuel Silva$^{1}$, Sarah Mancina$^{1,2}$, Jesse Osborn$^{1*}$\\
{$^{1}$ \itshape University of Wisconsin, Madison}\\
{$^{2}$ \itshape Universit{\`a} Degli Studi di Padova}\\
$^*$ Presenter

\ConferenceLogo{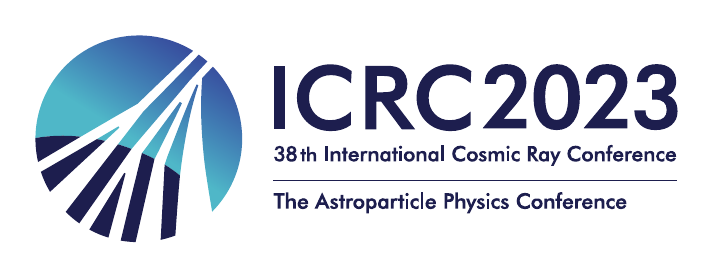}

\FullConference{The 38th International Cosmic Ray Conference (ICRC2023)\\ 26 July -- 3 August, 2023\\ Nagoya, Japan}
}

\begin{document}
\maketitle

\section{Introduction}\label{Intro}
High energy neutrinos of an astrophysical origin were discovered by IceCube in 2013 \cite{IceCube:2013low,PhysRevLett.111.021103, IceCube:2014stg}. Since then, IceCube searches for sources of these neutrinos suggest a non-negligible flux in the direction of TXS 0506+056 \cite{IceCube:2018cha,IceCube:2018dnn}, NGC 1068 \cite{NGC1068}, and from Milky Way galactic plane emission \cite{IceCube:DNNCSci,dnnc:2023icrc}. However, the flux seen from the sum of these sources is not sufficient to account for the total observed diffuse neutrino flux. Therefore, it is of great importance to measure the total astrophysical diffuse flux to infer properties of the production mechanisms of these neutrinos in a more model-independent manner. Historically, this is done assuming a single power law flux.

We first present a new data sample which relies on veto techniques and machine learning classification to reduce the muon rate in the southern sky while retaining a large sample of starting track events. These veto-based datasets also benefit from the self-veto effect \cite{PhysRevD.79.043009,PhysRevD.90.023009,NuVeto} which is a suppression of the atmospheric neutrino rate in the southern sky due to the removal of atmospheric muons from the same shower. Starting track events occur when a muon neutrino undergoes a charged current deep inelastic scattering interaction within the fiducial volume of the detector. A large initial cascade is observed from the hadronic component of the interaction, which results in a neutrino energy resolution of $\sim25-30\%$ between 1 TeV and 10 PeV. In addition, a muon track is produced in the interaction which is then used to reconstruct the neutrino direction. The resulting directional resolution is $1.6^\circ$ at 1 TeV and improves to $0.66^\circ$ at 100 TeV neutrino energies. 

We then present a measurement of the astrophysical diffuse flux using 10.3 years of data. This measurement is first under the single power law flux assumption with a 90\% sensitive energy range of 3 to 550 TeV. This is followed up with a more detailed measurement under the broken power law flux assumption. We also measure the flux as a sum of 8 single power law segments with a spectral index of $\gamma=2$. We then present a search of astrophysical neutrino sources, in particular in the direction of the Milky Way galaxy under the $Fermi$-LAT $\pi^0$~\cite{Fermi:FermiPi0Model} model assumption. Lastly, we estimate the contribution of these neutrinos from the galactic plane to the astrophysical diffuse flux.

\section{Event Selection}\label{ESTES}

\begin{figure}[t]
\centering
\begin{minipage}{.45\textwidth}
  \centering
  \includegraphics[width=1\linewidth]{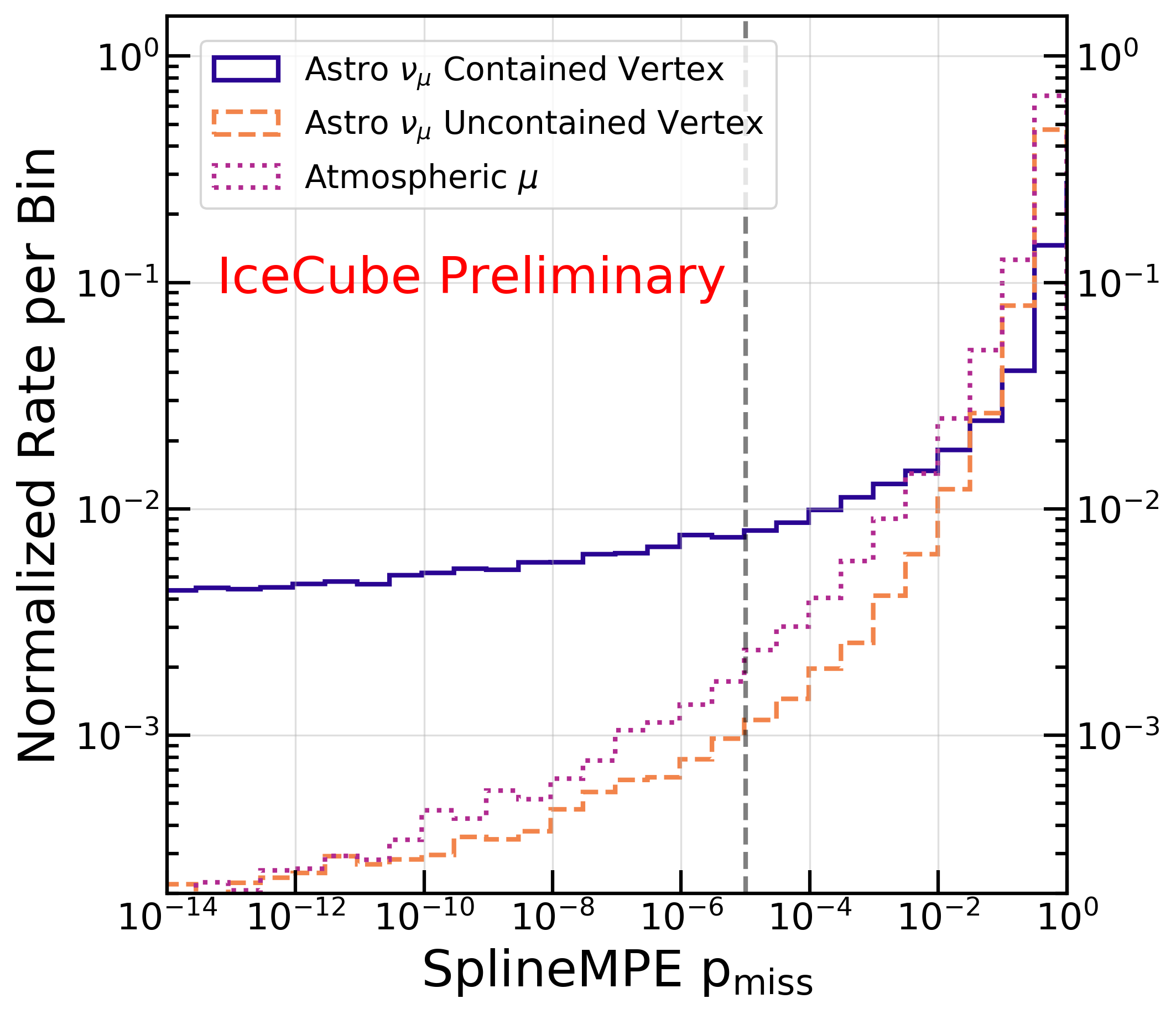}
  \captionof{figure}{p$_{\mathrm{miss}}$ distribution for atmospheric muons and astrophysical neutrinos. A cut of p$_{\mathrm{miss}} < 10^{-5}$ reduces atmospheric muon rates by three orders of magnitude. This figure is normalized to show p$_{\mathrm{miss}}$ shape differences between the interaction types.}
  \label{fig:pmiss}
\end{minipage}%
\hfill
\begin{minipage}{.45\textwidth}
  \includegraphics[width=1\linewidth]{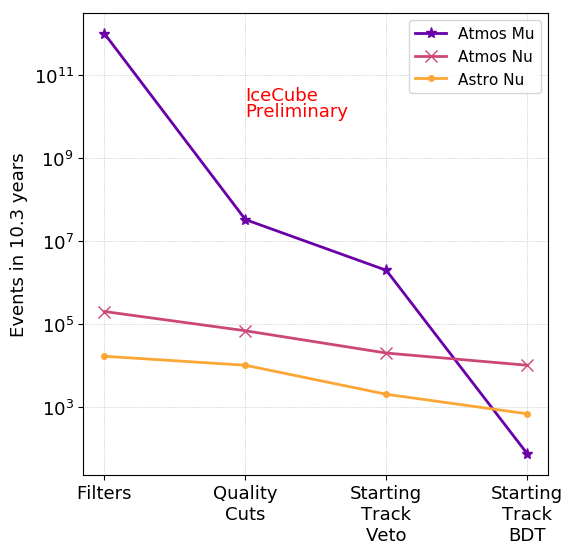}
  \centering
  \captionof{figure}{The atmospheric muon rate starts >5 orders of magnitude higher than the neutrino rate. After applying all the cuts, the neutrinos outnumber the muons by 2 orders of magnitude. The MC expectation shown is using benchmark models and shown for illustrative purposes only.}
  \label{fig:cutssummary}
\end{minipage}
\end{figure}

The Enhanced Starting Track Event Selection (ESTES) \cite{Silva:2019fnq,Mancina:2019hsp,IceCube:2021ctg,Aartsen:2023prd:estes} is a data sample within IceCube searching for starting track events. To reduce the overwhelming atmospheric muon background, the starting-track-veto (STV) was applied. The STV constructs a dynamic veto region using the properties of the muon track to determine how ``starting-like" the individual event is. The probability of an event being an incoming muon is referred to as the $\mathrm{p}_\mathrm{miss}$ and shown in Fig. \ref{fig:pmiss}. We note the atmospheric muons are reduced by about three orders magnitude with a small reduction in starting astrophysical neutrinos. The next significant cut is referred to as the ``starting track BDT". We use the XGBoost classifier \cite{Chen:2016:XST:2939672.2939785} to distinguish atmospheric muons against starting muon neutrino events. A cut is applied on the BDT score to further reduce the atmospheric muon rate to about a handful of expected muons per year. Further description of the BDT can found in references \cite{Silva:2019fnq,IceCube:2021ctg}. A summary of the cuts applied in ESTES are shown in Fig. \ref{fig:cutssummary}. We see a significant drop in the atmospheric muons while still retaining a significant number of neutrinos.

\section{Astrophysical Diffuse Flux Measurement}\label{Measurement}
\begin{wrapfigure}[8]{r}{.5\textwidth}
\begin{equation}
\begin{split}
&\Phi_\mathrm{Astro}^\mathrm{Total} = \phi_{\mathrm{Astro}}^{\mathrm{per-flavor}} \times (\frac{\mathrm{E}_{\nu}}{100 \mathrm{TeV}})^{-\gamma_{\mathrm{Astro}}} \times \mathrm{C}_{0}, \\
&\mathrm{C}_{0} = 3 \times 10^{-18} \times \mathrm{GeV}^{-1} \mathrm{cm}^{-2} \mathrm{s}^{-1}  \mathrm{sr}^{-1} \\
&\mathrm{where},\\
&\phi_{\mathrm{Astro}}^{\mathrm{per-flavor}} = 1.68 ^{+0.19}_{-0.22}, \quad \gamma_\mathrm{Astro} = 2.58 ^{+0.10}_{-0.09}\\
\label{eq:SPL}
\end{split}
\end{equation}
\end{wrapfigure}

\begin{figure}[h]
\centering
\begin{minipage}{.45\textwidth}
  \centering
  \includegraphics[width=1\linewidth]{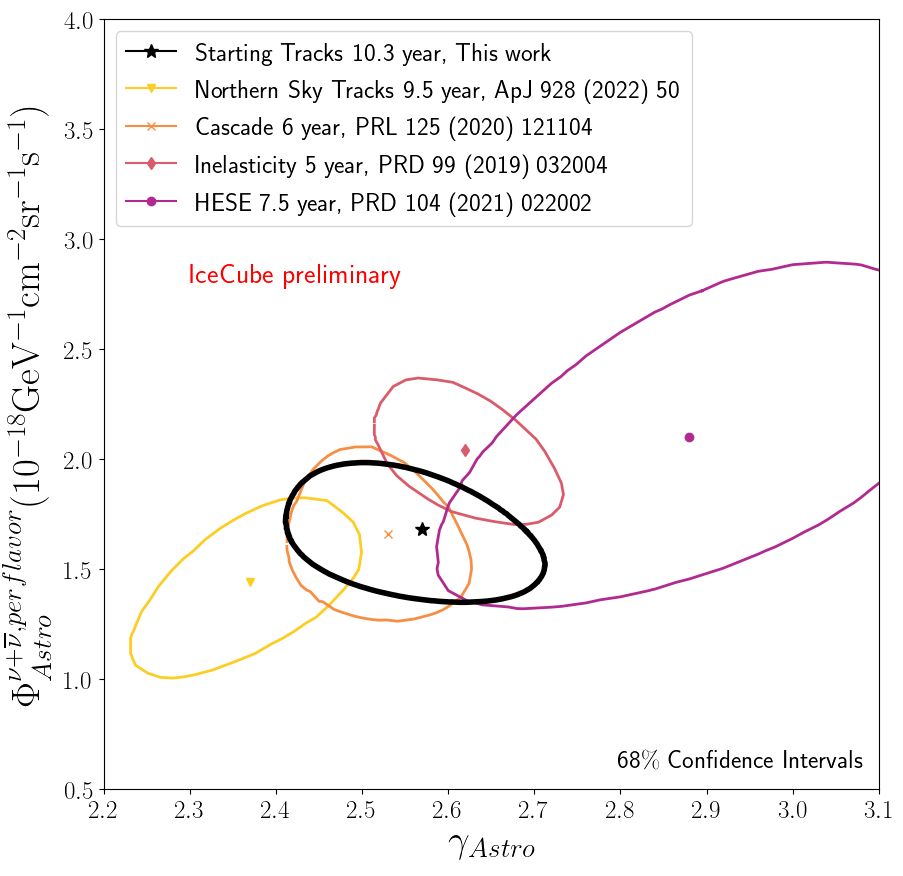}
  \captionof{figure}{A summary of the SPL 68\% confidence intervals for the Starting Tracks 10.3 year (this work) SPL measurement. We include recent IceCube results for direct comparison \cite{IceCube:2021uhz,SBUCasc,inelasticity,HESENew}. With these confidence intervals, all IceCube publications are consistent with this new result.}
  \label{fig:splfluxsummary}
\end{minipage}%
\hfill
\begin{minipage}{.45\textwidth}
  \includegraphics[width=1\linewidth]{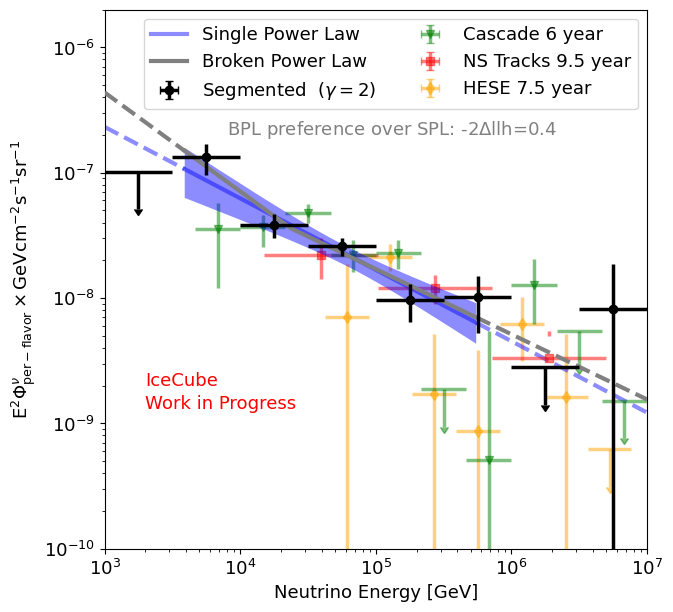}
  \centering
  \captionof{figure}{The blue line with error-band corresponds to the SPL measurement shown in Fig. \ref{fig:splfluxsummary}. The shaded region is the 90\% sensitive energy range computed using the SPL. The gray line is a fit to data assuming a broken power law flux. This model is not preferred in the data. The black points are the segmented power law flux measurement. We include results from recent IceCube publications for direct comparison \cite{SBUCasc,IceCube:2021uhz,HESENew}.}
  \label{fig:fluxsummary}
\end{minipage}
\end{figure}

A measurement of the astrophysical neutrino flux is performed using an isotropic single power law flux hypothesis (SPL) as shown in Eq. \ref{eq:SPL}. This model assumes $\nu_{e}:\nu_{\mu}:\nu_{\tau} = 1:1:1$ and $\nu:\bar\nu=1:1$ arriving at the surface of the Earth. In Eq. \ref{eq:SPL}, $\phi_{\mathrm{Astro}}^{\mathrm{per-flavor}}$ refers to the per flavor normalization scaling factor and is defined as a unit-less number. We include a treatment of the conventional and prompt atmospheric neutrino flux and the atmospheric muons. This measurement is inclusive of detector systematics better described in an upcoming publication \cite{Aartsen:2023prd:estes}. We measure a spectral index of $\gamma_{\mathrm{Astro}} = 2.58 ^{+0.10}_{-0.09}$ and per-flavor normalization of $\phi_{\mathrm{Astro}}^{\mathrm{per\ flavor}} = 1.68 ^{+0.19}_{-0.22}$ when measured at 100 TeV. The two-dimensional 68\% confidence intervals (CI) for the two parameters of interest are shown in Fig. \ref{fig:splfluxsummary} using the profile likelihood assuming Wilks' theorem \cite{Wilks}. Within the 68\% CI, all IceCube publications are consistent. Using the SPL, we then measure the 90\% sensitive energy as 3-550 TeV making this measurement the lowest energy measurement to-date. We then measure the astrophysical diffuse flux as a sum of 8 segments (assuming a single power law with $\gamma=2$) with results shown in Fig.  \ref{fig:fluxsummary} as black markers. All segments are consistent with a single power law.

\begin{figure}[h]
\centering
\includegraphics[width=0.45\textwidth]{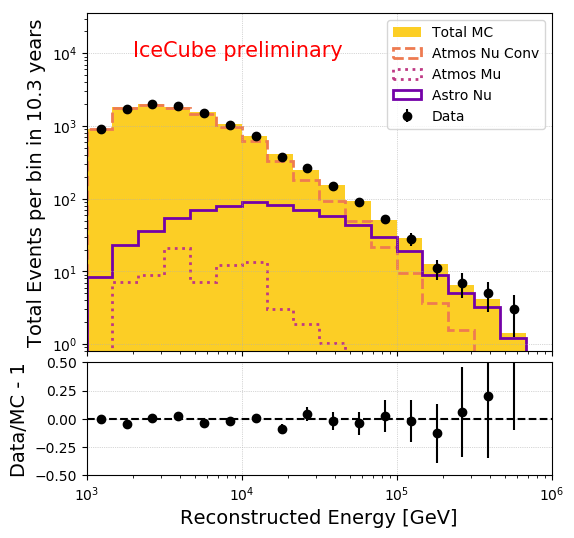}
\includegraphics[width=0.45\textwidth]{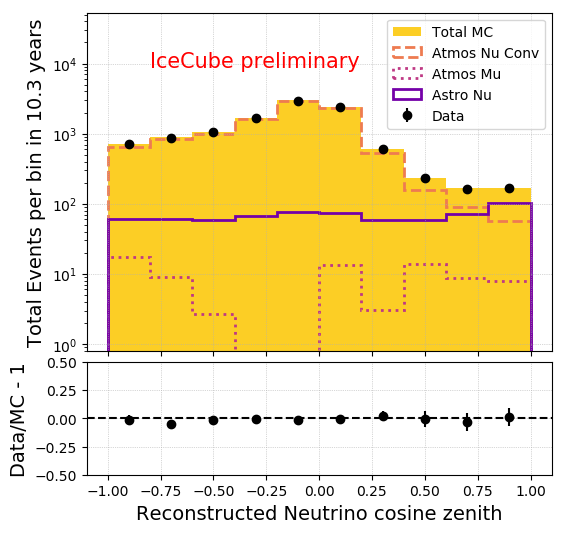}
\caption{The reconstructed energy and cosine zenith distributions for data and Monte Carlo using the best-fit parameters from the single power law flux measurement. The astrophysical neutrinos are shown as a solid purple line, the atmospheric neutrino and muon expectations are shown as dashed and dotted lines respectively. (The prompt flux is not shown since the best-fit is 0.)}
\label{fig:datamc}
\end{figure}

Using the SPL flux measurement, we now show the best-fit Monte Carlo expectation and compare it to 10.3 years of data in Fig. \ref{fig:datamc}. We observe excellent agreement with the data for all energy ranges and zenith angles of interest. In particular, the zenith distribution shows the atmospheric muons rates greatly reduced with little reduction in the astrophysical neutrino rates. A tabular summary of the number of events is shown in Tab. \ref{tab:eventcounts} with and without a cut applied in the southern sky.

\begin{wrapfigure}[11]{r}{.5\textwidth}
\begin{equation}
\begin{split}
&\Phi_{Astro}^{Total} = \phi_{0} \times (\frac{\mathrm{E}_{\mathrm{break}}}{100 \mathrm{TeV}})^{-\gamma_2} \times \mathrm{C}_{0},
\\
&\phi_{0} = 
\left\{
    \begin{array}{c}
        \phi^{\mathrm{Astro}}_{\mathrm{per-flavor}}  \times (\frac{\mathrm{E}}{\mathrm{E}_{\mathrm{break}}})^{-\gamma_1} (E < E_{\mathrm{break}})\\
        \phi^{\mathrm{Astro}}_{\mathrm{per-flavor}}  \times (\frac{\mathrm{E}}{\mathrm{E}_{\mathrm{break}}})^{-\gamma_2} (E > E_{\mathrm{break}})
    \end{array} 
\right. \space 
\\
&\mathrm{where},\\
&\phi^{\mathrm{Astro}}_{\mathrm{per-flavor}} = 1.7 ^{+0.19}_{-0.22}, \quad log_{10}(\mathrm{E}_\mathrm{break}) \sim 4.36 \\
&\gamma_{1} = 2.79 ^{+0.30}_{-0.50}, \quad \gamma_{2} = 2.52 ^{+0.10}_{-0.09}\\
\end{split}
\label{eq:bpl}
\end{equation}
\end{wrapfigure}


Given the sensitivity to sub-10 TeV energies, we then attempt to measure the astrophysical diffuse flux using a broken power law (BPL) shown analytically in Eq. \ref{eq:bpl}. The BPL is modeled as two power laws at a distinct energy break; one spectral index for energies below and one above this break. This gives us a model-independent way to probe for structure in the neutrino flux towards lower energies. However, the BPL is found to be consistent with the SPL with a test-statistic of 0.4. As a result, $E_{break}$ is poorly constrained, so we quote only the best fit point along with errors on $\gamma_{1}$ and $\gamma_{2}$ at that fixed $E_{break}$ from one-dimensional profile likelihood scans. This result is shown in Fig. \ref{fig:fluxsummary} as a gray line. 

This result is in mild tension with the preference for a break around this energy with an index $\gamma_{1}$= 1.31$^{+0.50}_{-1.21}$ presented in \cite{combinedfit:2023icrc}. We note that the sensitivity of the ESTES sample to an astrophysical flux at these energies arises predominantly from the southern sky (declinations $\delta$ < -35$^{\circ}$), while the analysis presented in \cite{combinedfit:2023icrc} is sensitive over the entire sky. Non-isotropic contributions, such as from the Galactic plane \cite{IceCube:DNNCSci} have not been considered in this broken power law hypothesis. In addition, systematic uncertainties have been parameterized differently in both analyses. Further investigations are ongoing about the potential impact of these parameterizations on the low-energy spectrum.

\begin{table}[h]
\centering
\begin{tabular}{||p{3cm}  | p{4cm} | p{4cm} ||} 
\hline
& Events with Zenith $< 80^\circ$ & All Events \\
\hline \hline
Astro Nu & 298 & 680 \\ 
Atmos Conv. Nu & 980 & 10042 \\ 
Atmos Conv. Mu & 42 & 75 \\ 
Total MC & 1320 & 10797 \\
\hline 
Data & 1365 & 10798 \\
\hline
\end{tabular}
\caption{A summary of observed data events and fitted MC events broken up by type under the single power law flux assumption. The left column has an additional cut on zenith $< 80^\circ$ to emphasize astrophysical neutrino purity in the southern sky.}
\label{tab:eventcounts}
\end{table}

\section{Neutrino Sources from the Galactic Plane}\label{GP}

\begin{figure}[h]
\centering
\includegraphics[width=0.9\textwidth]{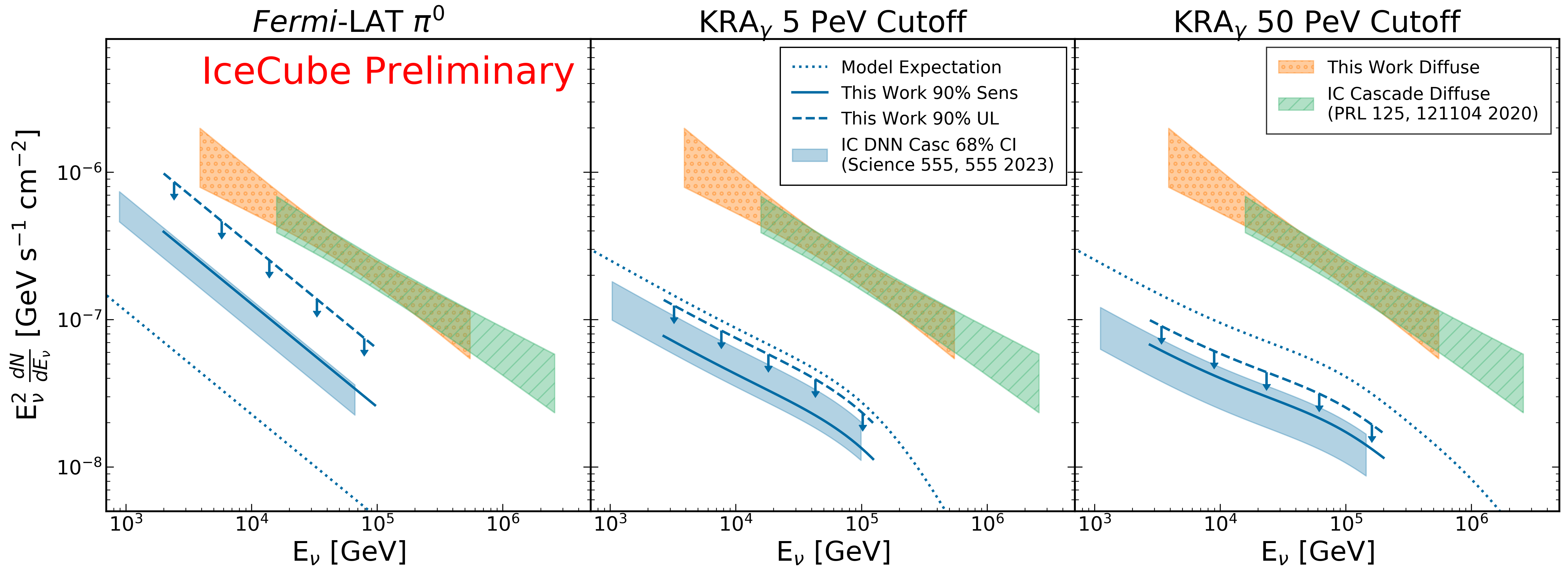}
\caption{The 90\% upper limits (blue dashed line) and 90\% sensitivities (solid blue line) for ESTES on the GP template analyses compared to the DNN cascade 68\% confidence interval (CI, blue shaded region) and the ESTES diffuse fit (orange shaded region) and Cascade diffuse fit (green shaded region). Each figure represents the diffuse galactic plane neutrino emission model tested: (Left) $Fermi$-LAT $\pi^0$~\cite{Fermi:FermiPi0Model}, (Center) KRA$_{\gamma}$ with a 5 PeV exponential energy cutoff, and (Right) KRA$_{\gamma}$ with a 50 PeV exponential energy cutoff~\cite{Gaggero:KRAgModel}  The energy range for this work's upper limit and sensitivity were calculated by injecting events according to the models spectral model and finding the central range where 90\% of the expected events lie.}
\label{fig:GPSummary}
\end{figure}

The purity of ESTES in the southern sky made it of interest for searching for astrophysical neutrino sources.
Four source hypotheses were tested including a search for neutrinos from the diffuse galactic plane (GP) emission.
Under the diffuse galactic plane hypothesis, the same cosmic ray flux seen at earth is assumed to be propagating throughout our galaxy and produces neutrinos through pions generated when the cosmic rays interact with the galactic plane medium.
Some of the most dense parts of the galactic plane, including the galactic center, are found in the southern sky where this event selection can improve upon previous IceCube searches with muon neutrinos due to the atmospheric neutrino self-veto effect.

The results using only ESTES returned a post-trial p-value of 0.070 for the $Fermi$-LAT $\pi^0$ map~\cite{Fermi:FermiPi0Model}, so we were unable to reject the null hypothesis.
The search also looked for neutrinos correlated with the locations of three galactic plane source catalogs: supernova remnants, pulsar wind nebulae, and unidentified TeV galactic plane objects. The supernova remnants resulted in the largest test statistic and the post-trial p-value was found to be 0.056, meaning the null hypothesis could not be rejected. Figure~\ref{fig:GPSummary} shows the 90\% upper limits set by this analysis on the three galactic plane templates tested in comparison to the best fits produced by IceCube's DNN-based cascade event selection~\cite{IceCube:DNNCSci}. These upper limits are consistent with \cite{IceCube:DNNCSci} and the sensitivity of this analysis.

\begin{wrapfigure}[16]{rt}{0.5\textwidth}
\vspace{-0.5em}
\centering
\includegraphics[width=0.5\textwidth]{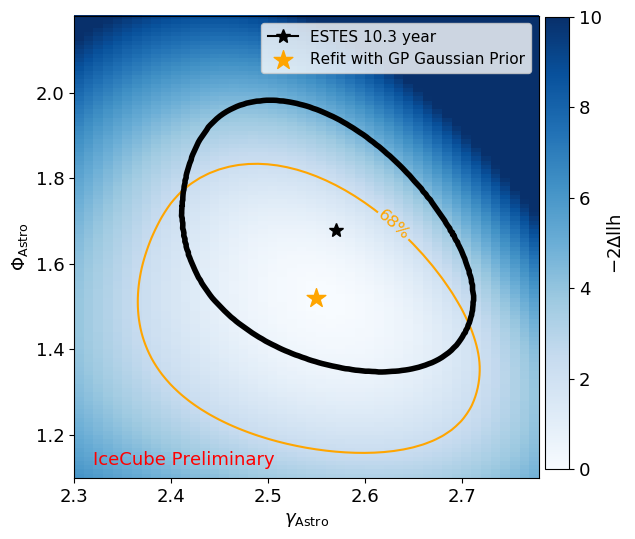}
\caption{Single power law flux measurement as described in the primary text of this work (in black) and introducing the galactic plane as additional nuisance parameter in the fit (in yellow).}
\label{fig:IsotropicFluxwGP}
\end{wrapfigure}

We then use the GP flux from the DNN-based cascade sample~\cite{IceCube:DNNCSci} and include it as an additional parameter (with a Gaussian constraint) in the measurement of the diffuse flux from Sec. \ref{Measurement}. Figure~\ref{fig:IsotropicFluxwGP} shows the measurement performed without the galactic plane neutrinos (in black) and now included (in yellow) using the $Fermi$-LAT $\pi^0$ template. We observe a large shift in the astrophysical normalization of $10\%$ with a negligible shift in the spectral index ($\Delta \gamma = 0.01$). This observation is in agreement with an independent measurement of the diffuse flux using through-going tracks from the northern sky \cite{gpnstracks:2023icrc}. Therefore, it is recommended that future isotropic diffuse flux measurements include the galactic plane as was shown in this work.

\section{Conclusion}\label{Conclusion}

A new dataset using veto-techniques and boosted decision trees to search for starting track events in the northern and southern hemisphere was shown. We detected 10,798 events over the entire sky with $99.3\%$ neutrino purity. This dataset found 1000 neutrinos of astrophysical origin with 1/3 of the astrophysical neutrinos localized to the southern sky.

This dataset was then used to measure the astrophysical neutrino flux using a single power law flux. The best-fit spectral index is $\gamma_\mathrm{Astro} = 2.58 ^{+0.10}_{-0.09}$ and per-flavor normalization is $\Phi_{\mathrm{Astro}}^{\mathrm{per-flavor}} = 1.68 ^{+0.19}_{-0.22}$ (at 100\,TeV). 
The sensitive energy ranges for this particular flux model was defined as 3-550\,TeV demonstrating the IceCube detector's ability to investigate the astrophysical neutrino flux with such precision in energy regions typically dominated by atmospheric backgrounds, below 16 TeV, for the first time. 
Assuming a single power law flux, we then presented a segmented measurement of the flux from 1 TeV to 10 PeV showing consistent segments with the single power law. Finally, a measurement of the flux under the broken power law assumption was performed. We tested a lower (higher) energy spectral index below (above) a break energy. We measured the BPL to be consistent with the SPL with a test-statistic of 0.4. 

We observe consistency above 10 TeV for the SPL parameters with all other IceCube samples, including the latest, statistically independent, combined measurement \cite{combinedfit:2023icrc}. Given that this analysis and \cite{combinedfit:2023icrc} have different systematic uncertainties and use different event selections and reconstruction techniques, the flux below 10 TeV may be impacted by additional systematic effects, which require further study.

A time-integrated neutrino source search was preformed searching for galactic plane diffuse neutrino emission. No significant excesses are found; however, important limits are set on the hadronic emission from TeV gamma-ray galactic plane objects. The upper limits from the galactic plane template analysis are consistent with the recent results using IceCube cascade events to study diffuse neutrino emissions from cosmic ray interactions in the galactic plane. However, there are efforts within IceCube to combine starting tracks with the other data streams to fully comprehend the galactic plane neutrino emission using all of the morphological structures produced by neutrinos within IceCube~\cite{IceCube:PPProceedings}.

Finally, we estimate the impact of the galactic plane neutrino emission under the Fermi $\pi^0$ flux model and concluded that while the expected impact on the diffuse flux spectral index is at the sub-percent level it can still contribute to the overall diffuse flux normalization by $\sim 10\%$. 

\bibliographystyle{ICRC}
\bibliography{references}

\providecommand{\href}[2]{#2}\begingroup\raggedright\begin{thebibliography}{10}

\bibitem{IceCube:2013low}
{\bfseries IceCube} Collaboration, M.~G. Aartsen {\em et~al.}
  \href{http://dx.doi.org/10.1126/science.1242856}{{\em Science} {\bfseries
  342} (2013) 1242856}.

\bibitem{PhysRevLett.111.021103}
{\bfseries IceCube} Collaboration, M.~G. Aartsen {\em et~al.}
  \href{http://dx.doi.org/10.1103/PhysRevLett.111.021103}{{\em Phys. Rev.
  Lett.} {\bfseries 111} (Jul, 2013) 021103}.

\bibitem{IceCube:2014stg}
{\bfseries IceCube} Collaboration, M.~G. Aartsen {\em et~al.}
  \href{http://dx.doi.org/10.1103/PhysRevLett.113.101101}{{\em Phys. Rev.
  Lett.} {\bfseries 113} (2014) 101101}.

\bibitem{IceCube:2018cha}
{\bfseries IceCube} Collaboration, M.~G. Aartsen {\em et~al.}
  \href{http://dx.doi.org/10.1126/science.aat2890}{{\em Science} {\bfseries
  361} no.~6398, (2018) 147--151}.

\bibitem{IceCube:2018dnn}
{\bfseries IceCube, Fermi-LAT, MAGIC, AGILE, ASAS-SN, HAWC, H.E.S.S., INTEGRAL,
  Kanata, Kiso, Kapteyn, Liverpool Telescope, Subaru, Swift NuSTAR, VERITAS,
  VLA/17B-403} Collaboration, M.~G. Aartsen {\em et~al.}
  \href{http://dx.doi.org/10.1126/science.aat1378}{{\em Science} {\bfseries
  361} no.~6398, (2018) eaat1378}.

\bibitem{NGC1068}
{\bfseries IceCube} Collaboration, R.~Abbasi {\em et~al.}
  \href{http://dx.doi.org/10.1126/science.abg3395}{{\em Science} {\bfseries
  378} no.~6619, (2022) 538--543}.

\bibitem{IceCube:DNNCSci}
{\bfseries IceCube} Collaboration, R.~Abbasi {\em et~al.}
  \href{http://dx.doi.org/10.1126/science.adc9818}{{\em Science} {\bfseries
  555} no.~5555, (2023) 555--555}.

\bibitem{dnnc:2023icrc}
{\bfseries IceCube} Collaboration, S.~Sclafani and M.~Hünnefeld {\em PoS}
  {\bfseries ICRC2023} (these proceedings) 1108.

\bibitem{PhysRevD.79.043009}
S.~Sch\"onert, T.~K. Gaisser, E.~Resconi, and O.~Schulz
  \href{http://dx.doi.org/10.1103/PhysRevD.79.043009}{{\em Phys. Rev. D}
  {\bfseries 79} (Feb, 2009) 043009}.

\bibitem{PhysRevD.90.023009}
T.~K. Gaisser, K.~Jero, A.~Karle, and J.~van Santen
  \href{http://dx.doi.org/10.1103/PhysRevD.90.023009}{{\em Phys. Rev. D}
  {\bfseries 90} (Jul, 2014) 023009}.

\bibitem{NuVeto}
C.~A. Arg\"{u}elles, S.~Palomares-Ruiz, A.~Schneider, L.~Wille, and T.~Yuan
  \href{http://dx.doi.org/10.1088/1475-7516/2018/07/047}{{\em Journal of
  Cosmology and Astroparticle Physics} {\bfseries 2018} no.~07, (Jul, 2018)
  047--047}.

\bibitem{Fermi:FermiPi0Model}
{\bfseries Fermi-LAT} Collaboration, M.~Ackermann {\em et~al.}
  \href{http://dx.doi.org/10.1088/0004-637X/750/1/3}{{\em Astrophys. J.}
  {\bfseries 750} (2012) 3}.

\bibitem{Silva:2019fnq}
{\bfseries IceCube} Collaboration, M.~Silva and S.~Mancina
  \href{http://dx.doi.org/10.22323/1.358.1010}{{\em PoS} {\bfseries ICRC2019}
  (2020) 1010}.

\bibitem{Mancina:2019hsp}
{\bfseries IceCube} Collaboration, S.~Mancina and M.~Silva
  \href{http://dx.doi.org/10.22323/1.358.0954}{{\em PoS} {\bfseries ICRC2019}
  (2020) 954}.

\bibitem{IceCube:2021ctg}
{\bfseries IceCube} Collaboration, R.~Abbasi {\em et~al.}
  \href{http://dx.doi.org/10.22323/1.395.1130}{{\em PoS} {\bfseries ICRC2021}
  (2021) 1130}.

\bibitem{Aartsen:2023prd:estes}
{\bfseries IceCube} Collaboration, M.~G. Aartsen {\em et~al.} {\em To be
  submitted to Physical Review D} .

\bibitem{Chen:2016:XST:2939672.2939785}
T.~Chen and C.~Guestrin,
  \href{http://dx.doi.org/10.1145/2939672.2939785}{``{XGBoost}: A scalable tree
  boosting system,''} in {\em Proceedings of the 22nd ACM SIGKDD International
  Conference on Knowledge Discovery and Data Mining}, KDD '16, pp.~785--794.
\newblock ACM, New York, NY, USA, 2016.
\newblock \url{http://doi.acm.org/10.1145/2939672.2939785}.

\bibitem{Wilks}
S.~S. Wilks \href{http://dx.doi.org/10.1214/aoms/1177732360}{{\em The Annals of
  Mathematical Statistics} {\bfseries 9} no.~1, (1938) 60 -- 62}.

\bibitem{combinedfit:2023icrc}
{\bfseries IceCube} Collaboration, R.~Naab, E.~Ganster, and Z.~Zhang {\em PoS}
  {\bfseries ICRC2023} (these proceedings) 1064.

\bibitem{IceCube:2021uhz}
{\bfseries IceCube} Collaboration, R.~Abbasi {\em et~al.}
  \href{http://dx.doi.org/10.3847/1538-4357/ac4d29}{{\em Astrophys. J.}
  {\bfseries 928} no.~1, (2022) 50}.

\bibitem{SBUCasc}
{\bfseries IceCube} Collaboration, M.~G. Aartsen {\em et~al.}
  \href{http://dx.doi.org/10.1103/PhysRevLett.125.121104}{{\em Phys. Rev.
  Lett.} {\bfseries 125} (Sep, 2020) 121104}.

\bibitem{inelasticity}
{\bfseries IceCube} Collaboration, M.~G. Aartsen {\em et~al.}
  \href{http://dx.doi.org/10.1103/PhysRevD.99.032004}{{\em Phys. Rev. D}
  {\bfseries 99} no.~3, (2019) 032004}.

\bibitem{HESENew}
{\bfseries IceCube} Collaboration, R.~Abbasi {\em et~al.}
  \href{http://dx.doi.org/10.1103/PhysRevD.104.022002}{{\em Phys. Rev. D}
  {\bfseries 104} (Jul, 2021) 022002}.

\bibitem{Gaggero:KRAgModel}
D.~Gaggero, D.~Grasso, A.~Marinelli, A.~Urbano, and M.~Valli.

\bibitem{gpnstracks:2023icrc}
{\bfseries IceCube} Collaboration, P.~Fuerst {\em PoS} {\bfseries ICRC2023}
  (these proceedings) 1046.

\bibitem{IceCube:PPProceedings}
{\bfseries IceCube} Collaboration, P.~Savina, L.~Lu, and T.~Yuan {\em PoS}
  {\bfseries ICRC2023} (these proceedings) 1010.

\end{thebibliography}\endgroup

\clearpage
\section*{Full Author List: IceCube Collaboration}

\scriptsize
\noindent
R. Abbasi$^{17}$,
M. Ackermann$^{63}$,
J. Adams$^{18}$,
S. K. Agarwalla$^{40,\: 64}$,
J. A. Aguilar$^{12}$,
M. Ahlers$^{22}$,
J.M. Alameddine$^{23}$,
N. M. Amin$^{44}$,
K. Andeen$^{42}$,
G. Anton$^{26}$
% ...
\\
\\
% ...
$^{12}$ Universit{\'e} Libre de Bruxelles, Science Faculty CP230, B-1050 Brussels, Belgium \\
% ...
$^{17}$ Department of Physics, Loyola University Chicago, Chicago, IL 60660, USA \\
$^{18}$ Dept. of Physics and Astronomy, University of Canterbury, Private Bag 4800, Christchurch, New Zealand \\
% ...
$^{22}$ Niels Bohr Institute, University of Copenhagen, DK-2100 Copenhagen, Denmark \\
$^{23}$ Dept. of Physics, TU Dortmund University, D-44221 Dortmund, Germany \\
% ...
$^{26}$ Erlangen Centre for Astroparticle Physics, Friedrich-Alexander-Universit{\"a}t Erlangen-N{\"u}rnberg, D-91058 Erlangen, Germany \\
% ...
$^{40}$ Dept. of Physics and Wisconsin IceCube Particle Astrophysics Center, University of Wisconsin{\textendash}Madison, Madison, WI 53706, USA \\
% ...
$^{42}$ Department of Physics, Marquette University, Milwaukee, WI, 53201, USA \\
% ...
$^{44}$ Bartol Research Institute and Dept. of Physics and Astronomy, University of Delaware, Newark, DE 19716, USA \\
% ...
$^{63}$ Deutsches Elektronen-Synchrotron DESY, Platanenallee 6, 15738 Zeuthen, Germany  \\
$^{64}$ Institute of Physics, Sachivalaya Marg, Sainik School Post, Bhubaneswar 751005, India \\
% ...

\subsection*{Acknowledgements}

\noindent
The authors gratefully acknowledge the support from the following agencies and institutions:
USA {\textendash} U.S. National Science Foundation-Office of Polar Programs,
U.S. National Science Foundation-Physics Division,
U.S. National Science Foundation-EPSCoR,
Wisconsin Alumni Research Foundation,
Center for High Throughput Computing (CHTC) at the University of Wisconsin{\textendash}Madison,
Open Science Grid (OSG),
Advanced Cyberinfrastructure Coordination Ecosystem: Services {\&} Support (ACCESS),
Frontera computing project at the Texas Advanced Computing Center,
U.S. Department of Energy-National Energy Research Scientific Computing Center,
Particle astrophysics research computing center at the University of Maryland,
Institute for Cyber-Enabled Research at Michigan State University,
and Astroparticle physics computational facility at Marquette University;
Belgium {\textendash} Funds for Scientific Research (FRS-FNRS and FWO),
FWO Odysseus and Big Science programmes,
and Belgian Federal Science Policy Office (Belspo);
Germany {\textendash} Bundesministerium f{\"u}r Bildung und Forschung (BMBF),
Deutsche Forschungsgemeinschaft (DFG),
Helmholtz Alliance for Astroparticle Physics (HAP),
Initiative and Networking Fund of the Helmholtz Association,
Deutsches Elektronen Synchrotron (DESY),
and High Performance Computing cluster of the RWTH Aachen;
Sweden {\textendash} Swedish Research Council,
Swedish Polar Research Secretariat,
Swedish National Infrastructure for Computing (SNIC),
and Knut and Alice Wallenberg Foundation;
European Union {\textendash} EGI Advanced Computing for research;
Australia {\textendash} Australian Research Council;
Canada {\textendash} Natural Sciences and Engineering Research Council of Canada,
Calcul Qu{\'e}bec, Compute Ontario, Canada Foundation for Innovation, WestGrid, and Compute Canada;
Denmark {\textendash} Villum Fonden, Carlsberg Foundation, and European Commission;
New Zealand {\textendash} Marsden Fund;
Japan {\textendash} Japan Society for Promotion of Science (JSPS)
and Institute for Global Prominent Research (IGPR) of Chiba University;
Korea {\textendash} National Research Foundation of Korea (NRF);
Switzerland {\textendash} Swiss National Science Foundation (SNSF);
United Kingdom {\textendash} Department of Physics, University of Oxford.

\end{document}